\documentclass[aps,prb,10pt,amsmath,amssymb,twocolumn,footinbib,superscriptaddress]{revtex4-1}

\usepackage{color, braket}
\usepackage[large,bf]{subfigure}     

\usepackage[breaklinks]{hyperref}
\newcommand{\Z}[1]{\texorpdfstring{\ensuremath{\mathbb{Z}_{#1}}}{Z#1}}
\newcommand{\phd}{ {\vphantom{\dag}} }		
\newcommand{\hc}{ {\mathfrak{h.c.}} }

\definecolor{purple}{rgb}{0.5,0,0.5}
\definecolor{dkgreen}{rgb}{0,0.5,0}
\definecolor{orange}{rgb}{1,0.5,0}

\usepackage{ifpdf}
\ifpdf
    \usepackage{psfrag}
    \usepackage[pdftex]{epsfig}
\else
    \usepackage{psfrag}
    \usepackage[dvips]{epsfig}
\fi

\begin{document}

\title{Topological phases with parafermions: theory and blueprints}
\date{\today}
\author{Jason Alicea}
\affiliation{Department of Physics and Institute for Quantum Information and Matter, California Institute of Technology, Pasadena, CA 91125, USA}
\affiliation{Walter Burke Institute for Theoretical Physics, California Institute of Technology, Pasadena, CA 91125, USA}
\author{Paul Fendley}
\affiliation{All Souls College and Rudolf Peierls Centre for Theoretical Physics, 1 Keble Road, Oxford, OX1 3NP, UK}

\begin{abstract}
We concisely review the recent evolution in the study of parafermions---exotic emergent excitations that generalize Majorana fermions and similarly underpin a host of novel phenomena.  First we illustrate the intimate connection between $\Z{3}$-symmetric ``spin'' chains and one-dimensional parafermion lattice models, highlighting how the latter host a topological phase featuring protected edge zero modes.  We then tour several blueprints for the laboratory realization of parafermion zero modes---focusing on quantum Hall/superconductor hybrids, quantum Hall bilayers, and two-dimensional topological insulators---and describe striking experimental fingerprints that they provide.  Finally, we discuss how coupled parafermion arrays in quantum Hall architectures yield topological phases that potentially furnish hardware for a universal, intrinsically decoherence-free quantum computer.  
\end{abstract}

\maketitle

\section{Introduction}

The theoretical and experimental search for topologically nontrivial condensed-matter systems with fractionalized excitations 
continues to attract widespread attention.  One reason for the excitement is fundamental, as fractionalization brilliantly illustrates the rich emergent long-distance behavior that many-body systems can exhibit.  Another reason is more practical: certain `non-Abelian' fractionalized excitations form the bedrock of topological quantum computers that tantalizingly promise inherent immunity against errors \cite{Kitaev:2003,Nayak:2008}.

Although fractionalized topological phases appear quite exotic, several fundamental examples of such states share a profound relationship with the very well known Ising quantum spin chain.  One facet of this relationship arises from the Jordan-Wigner transformation, a non-local map between the bosonic Ising spins and fermions. While this mapping has long been known to provide a useful tool for solving the two-dimensional classical Ising model and its quantum-chain limit \cite{SML,McCoybook}, great insights arise by treating the fermions themselves as physical degrees of freedom.  With the latter viewpoint the quantum Ising chain corresponds to an electronic model for a one-dimensional (1D) $p$-wave superconductor.  

Remarkably, conventional spin ordering for the Ising chain yields a topological superconducting phase in fermionic language.\cite{Kitaev:2001} The degenerate `up' and `down' ordered spin configurations of the former translate into even- and odd-fermion-parity superconducting states. However, the degeneracy in the latter is resilient; a local magnetic field that readily splits the energy of the spin polarizations becomes non-local in fermionic language.  The $p$-wave superconductor thus exhibits a ground-state degeneracy immune to physical perturbations---one of the hallmarks of a topological phase.  

Because of their inherent robustness, the superconducting ground states can encode fault-tolerant `topological qubits'.  The qubits arise from Majorana zero modes that bind to the superconductor's edges.\cite{Kitaev:2001}  (Zero modes in the Ising spin chain with open boundary conditions were found long ago\cite{Pfeuty}, but their topological significance not appreciated until Kitaev's seminal work.)
More interestingly still, the superconductor supports the remarkable phenomenon of non-Abelian exchange statistics; that is, swapping pairs of zero modes 
rotates the system's quantum state---and hence the topological qubits---in exquisitely precise ways.\cite{Read:2000,Ivanov,AliceaBraiding}  These highly attractive properties have inspired enormous activity; recent reviews of this fast-moving field appear in Refs.~\onlinecite{BeenakkerReview,Alicea:2012n,FlensbergReview,TewariReview,FranzReview,ChetanReview}. 

The close connection between spin systems and topological phases extends in interesting and often surprising ways to `clock models' with $\Z{n}$ symmetry \cite{Fendley:2012}.  These models admit a non-local mapping to chains of {\em parafermions} that comprise $\Z{n}$ generalizations of Majorana operators\cite{FradkinKadanoff}.  Our goal here is to review the fantastically rich physics that has emerged through recent studies of these more exotic objects.  In many ways the story parallels the Ising/Majorana problem highlighted above, but certainly not trivially so; throughout we will encounter numerous features unique to parafermions.  

An immediately glaring distinction is that all of condensed matter arises from conventional bosons and fermions, so that explorations of parafermions might naively seem purely academic.  We stress that this is far from the case.  A substantial fraction of our discussion will survey plausible experimental routes to trapping parafermionic excitations even in well understood, presently available systems.  Although more challenging to implement, these platforms give rise to topological qubits that are both better protected against environmental noise and allow for richer fault-tolerant qubit rotations compared to Majorana-based architectures.  In fact we will eventually describe methods for leveraging systems with parafermions to synthesize fully universal topological quantum computing hardware!

\section{Parafermions in chains}

\label{sec:chains}

In this section we review the basics of spin chains with $\Z3$ symmetry and their description in terms of parafermions. While some of the analysis straightforwardly generalizes the $\Z2$ Ising/Majorana case, we highlight several facets not present there.  Our discussion both lays the groundwork for the novel experimental proposals discussed in subsequent sections and also illuminates several problems of more formal interest.

\subsection{Lattice parafermions}
\label{LatticeParafermionSec}

Clock models provide a natural $\Z{n}$-symmetric generalization of the Ising model. They exhibit similar ordered and disordered phases, and also admit a non-local representation wherein symmetry-breaking states map to topological phases supporting local zero modes\cite{Fendley:2012}. Additionally, for $n>2$ a variety of interesting behavior not present in the $n=2$ Ising case emerges, especially when the interactions are chiral \cite{Ostlund,Huse,Howes,Albertini}. The corresponding topological phases have been classified \cite{Motruk:2013,Bondesan:2013}.

We focus primarily on the 1D three-state clock model since it contains almost all the physics of interest. The Hilbert space consists of a three-state ``spin" on each site. The usual Pauli matrices $\sigma^z$ and $\sigma^x$ respectively generalize to $\sigma$ and $\tau$, defined by
\begin{equation}
\sigma = \begin{pmatrix} 1&& \\ &\omega& \\ &&\omega^2 \end{pmatrix}\ ,\qquad
\tau = \begin{pmatrix} &&1 \\ 1&& \\ &1& \end{pmatrix}\ , \qquad \omega = e^{i2\pi/3}.
\label{sigmataudef}
\end{equation}
Then the ``clock'' operator $\sigma_a = 1\otimes\dots\otimes 1\otimes \sigma \otimes 1\dots$ measures the spin on site $a$, while the ``shift'' operator $\tau_a= 1\otimes\dots\otimes 1\otimes \tau\otimes 1\dots$ winds its value by $\omega$.  These operators obey the algebra
\begin{equation}
		\sigma_a^3 = 1\ ,
	\qquad \tau_a^3 = 1\ ,
	\qquad \sigma_a \tau_a = \omega \tau_a \sigma_a
\end{equation}
and commute on different sites, e.g., $\sigma_a\tau_b = \tau_b\sigma_a$ for $a \neq b$.

The most general nearest-neighbor Hamiltonian for the three-state clock model on an open $L$-site chain is
\begin{align}
	H = -J \sum_{a=1}^{L-1} \big(e^{i\phi} \sigma^\dag_{a+1} \sigma_{a} + \hc \big)
		- f \sum_{a=1}^L \big( e^{i\theta} \tau^\dag_a + \hc )\ .
	\label{Hclock}
\end{align}
Throughout we assume purely real couplings $J, f, \phi$ and $\theta$, with $f$ and $J$ non-negative.  The Hamiltonian preserves several symmetries; see Ref.~\onlinecite{Mong:2014b} for a detailed discussion.  As a particularly important example, there is a global $\Z3$ symmetry corresponding to a uniform shift of all spins generated by $\mathcal{Q} = \prod_a \tau_a^\dag$, which satisfies $\mathcal{Q}^3=1$.  The Hamiltonian also possesses duality symmetries when $f=J$ and $\phi = \theta$ that strongly constrain the phase diagram.  

The ferromagnetic case $\phi=0\,$mod$\,2\pi/3$---usually called the three-state Potts model---has been exceptionally well-studied from many viewpoints\cite{Baxbook}; some recent efforts motivated by the connection to topological physics include Refs.~\onlinecite{Motruk:2013,Cheng:2014,Andrei}.  The model exhibits a self-dual critical phase transition at $f=J$ separating ordered ($f<J$) and disordered ($f>J$) phases.  In the ordered phase the system admits three degenerate ground states characterized by $\langle \sigma_a \rangle \neq 0$.  Acting with the generator $\mathcal{Q}$ cycles through these ground states, indicating spontaneous $\Z3$ symmetry breaking.  The disordered phase, by contrast, yields a unique, symmetric ground state with $\langle \sigma_a \rangle = 0$. 

When $\phi\ne 0\,$mod$\,\pi/3$, the clock model model is chiral, breaking spatial parity and time-reversal symmetries.  The chiral case exhibits a much more intricate phase structure than the ferromagnetic limit.  Some information can be obtained by exploiting integrability on the two-parameter manifold $J\cos(3\phi)=f\cos(3\theta)$, where for example an incommensurate phase with ground-state level crossings appears \cite{Albertini}.  A systematic numerical analysis was (finally!) performed recently, finding accurate phase boundaries between ordered, trivially disordered, and incommensurate phases\cite{Hughes:2015}. Interestingly, there is evidence that incommensurability occurs along the self-dual $f=J$, $\phi=\theta$ line all the way up to the ferromagnetic limit.  

Broken symmetry states of the clock model deeply relate to exotic topologically nontrivial phases of matter---a remarkable correspondence that becomes manifest upon exploiting the Fradkin-Kadanoff mapping\cite{FradkinKadanoff} that rewrites the Hamiltonian in terms of parafermions.   As with Majorana fermions in the Ising chain, parafermions are defined via products of `order operators' $\sigma_a$ and `disorder operators' built from $\tau$-strings that create kinks by winding a macroscopic set of spins. 
Precisely, one set of parafermion operators reads
\begin{eqnarray}
\alpha_{2a-1} &=& (\omega\sigma_a^\dag\tau_a^\dag) \tau_{a+1}^\dag \tau_{a+2}^\dag \cdots,
\nonumber \\
\alpha_{2a} &=& (\sigma_a^\dag) \tau_{a+1}^\dag \tau_{a+2}^\dag \cdots \ ,
\label{alphas}
\end{eqnarray}
which obey
\begin{equation}
\alpha_b^3=1\ ,\qquad \alpha_b\, \alpha_{c} = \omega^{ \operatorname{sgn}(c-b)}\, \alpha_c \,\alpha_{b} 
\ .
\label{paracomm}
\end{equation}
Notice that the parafermions do not commute at different sites; this important property reflects their non-locality in clock variables.  Equally importantly, they do not anticommute either, unlike fermions.  Instead exchanging parafermion operators yields a \emph{complex} phase of $\omega$---a situation reminiscent of anyons.

In terms of parafermions, the Hamiltonian is
\begin{align}
	H = -Je^{i\phi} \sum_{\!b\text{ even}\!} \omega^2 \alpha_{b+1}^\dag \alpha_{b}^\phd - fe^{i\theta} \sum_{b\text{ odd}} \omega^2 \alpha_{b+1}^\dag \alpha_{b}^\phd + \hc
	\label{Hpf}
\end{align}
Even though $H$ is bilinear in parafermions, the factor of $\omega^{\pm 1}$ in the algebra (\ref{paracomm}) means that the model cannot be solved simply by Fourier transformation (contrary to the Ising case).  It is worth noting however that the entire spectrum of the non-Hermitian Hamiltonian found by omitting `$+\hc$' in Eq.~(\ref{Hpf}) 
is in a precise sense that of ``free'' parafermions \cite{Baxter:1989,Fendley:2014}.

At this point an interesting question arises: exactly how is the physics of the ordered and disordered clock-model phases encoded in the parafermion representation?  The answer directly relates to the existence of parafermion zero modes, which play an essential role throughout this review.

\subsection{Edge zero modes}
\label{edgezero}

One important signal of a topological phase in 1D is the existence of robust edge zero modes, the analogue of gapless modes in higher dimensions.  To be precise, consider a system with a discrete symmetry generated by an operator $\mathcal{Q}$.  We will say that the system possesses a `strong zero mode' if there exists an operator that commutes with the microscopic Hamiltonian up to terms falling off exponentially with system size, but does not commute with $\mathcal Q$.  The energy levels for each eigenvalue of $\mathcal Q$ must then be identical (modulo exponentially small corrections), resulting in degeneracies across the entire spectrum.  Strong edge zero modes have the same property but additionally localize with exponential fall-off to the boundaries of an open system.  As a famous example, the topological phase in the Majorana-chain representation of the Ising model supports strong edge Majorana zero modes, which one can readily construct explicitly and exactly.\cite{Pfeuty,Kitaev:2001}

We emphasize that strong edge zero modes impose an extraordinarily strong constraint on the spectrum that propagates to arbitrarily high energies.  In glaring contrast, a 1D topological phase in general requires only ground-state degeneracy---a much weaker condition. It is thus useful to introduce the notion of a `weak edge zero mode' that only commutes with the Hamiltonian projected into a low-energy subspace, thus encoding the minimal degeneracy necessary for a topological phase.  While harder to find, strong zero modes do offer potential advantages, since one in principle could exploit the excited-state degeneracies to build a fault-tolerant qubit that works even at high temperatures.\cite{Huse:2013} 

\begin{figure}
\includegraphics[width=\columnwidth]{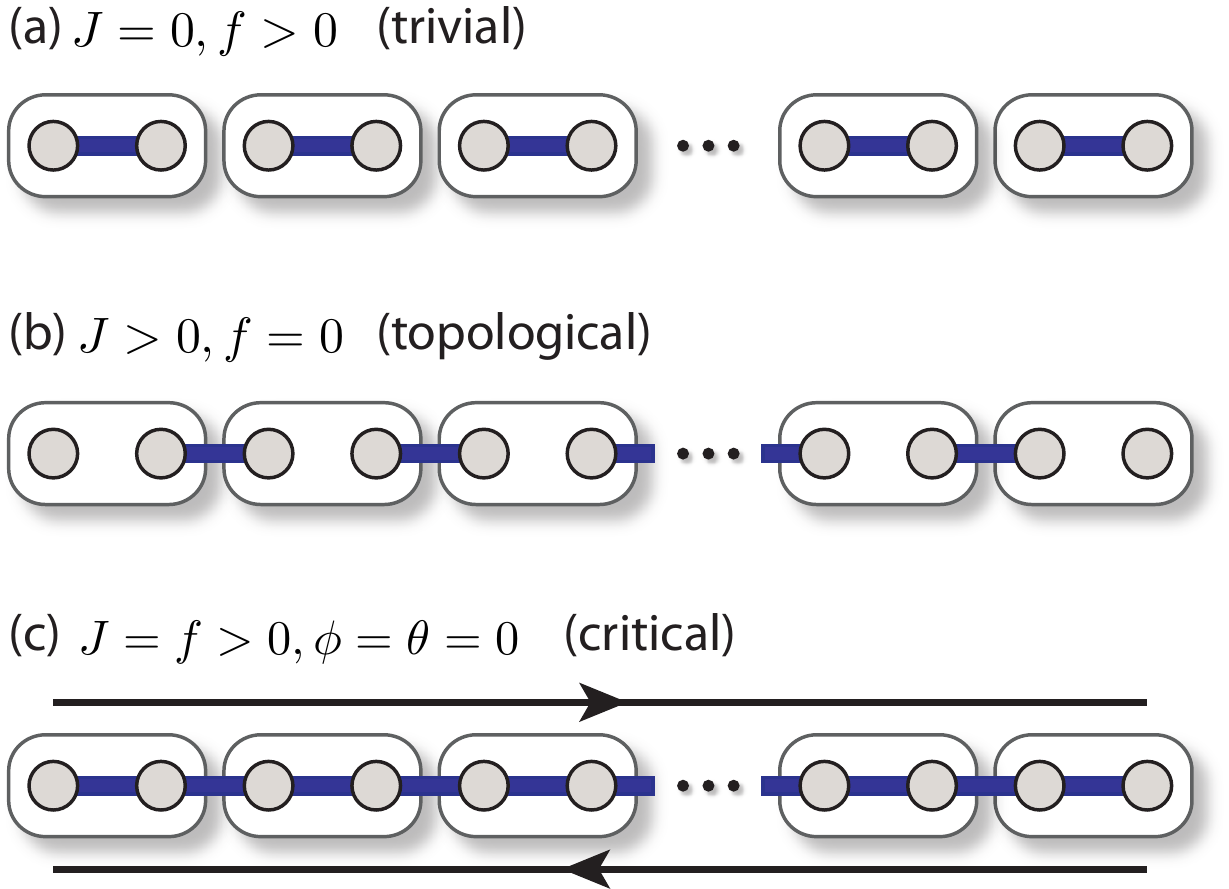}
\caption{Schematic representation of the parafermion chain [Eq.~(\ref{Hpf})] deep in the (a) trivial and (b) topological phases, and (c) at the intervening critical point. }
\label{Chain_fig}
\end{figure}

With these preliminaries in place, let us return to the parafermion representation of the clock chain.  
To establish a baseline we first examine two limits, illustrated in Figs.~\ref{Chain_fig}(a) and (b), where the model is trivially solvable.  In the figure 
circles within each box represent the two parafermions $\alpha_{2a-1}$ and $\alpha_{2a}$ associated with a single clock site $a$; recall Eqs.~(\ref{alphas}).  

In the maximally disordered limit $J=0$, spins on different sites simply decouple, and so parafermions interact only with their partner on the same clock site as indicated by the bonds in Fig.~\ref{Chain_fig}(a).  The parafermion chain then generically realizes a gapped trivial phase with a unique ground state, consistent with the physics in the spin representation.  

When $f=0$ the model maximally orders in spin language; accordingly, parafermions on adjacent clock sites strongly couple as represented by the bonds from one box to the next in Fig.~\ref{Chain_fig}(b).  Here too the bulk of the parafermion chain generically acquires a gap, though now the edges behave more interestingly.   In particular, a novel feature at $f=0$ is that the parafermions $\alpha_1$ and $\alpha_{2L}$ on the ends disappear from the Hamiltonian and thus commute with it.  They do not, however, commute with the $\mathbb{Z}_3$ symmetry generator $\mathcal{Q}$.  The system therefore exhibits a pair of \emph{strong} edge parafermion zero modes $\alpha_{1},\alpha_{2L}$ that signify the onset of a topological phase.  When acting on any mutual eigenstate of $H$ and $\mathcal Q$, the zero modes yield a distinct eigenstate with the same energy.  (The states must differ because if $q$ denotes the initial eigenvalue of $\mathcal{Q}$, acting with $\alpha_1$ or $\alpha_{2L}$ cycles the eigenvalue to $\omega q$.) Thus the ground states and all excited states assume at least threefold degeneracy in the $f = 0$ limit.  While the degeneracy has a topological origin in the parafermion representation, in spin language it arises from the three sectors of symmetry-breaking states with $\langle\sigma_a\rangle \neq 0$ and their excitations.  

The fate of the strong parafermion zero modes away from this trivial limit raises an important question:  are there modified operators $\alpha_{1} +\cdots$ and $\alpha_{2L}+\cdots$ that commute with the Hamiltonian even when $f \neq 0$?  A natural guess is that strong zero modes persist provided the bulk gap remains finite so that the topological phase is intact; this intuition indeed proves correct in the Ising/Majorana case.  
The interesting answer here, however, is that the strong zero modes require chirality  (i.e., $\phi \neq 0$ mod $\pi/3$) to survive!\cite{Fendley:2012}
This result is particularly surprising given that the topological phase is  most robust (i.e., has the largest gap) in the ferromagnetic limit and, more broadly, always remains for sufficiently small $f$ as long as the system orders ferromagnetically in spin language.\cite{Motruk:2013,Jermyn,Andrei}.
(Any topological significance of the incommensurate phase is a so-far unexplored issue.)  Instead, the topological phase in the non-chiral ferromagnetic $\phi = 0$ limit generically supports only \emph{weak} edge parafermion zero modes of the form $\alpha_{1,2L} + \cdots$ that nicely shift between three degenerate ground states but are not especially meaningful when acting on excited states.\cite{Andrei}

There are several ways of seeing why strong parafermion zero modes require chirality. 
One method utilizes an iterative procedure \cite{Fendley:2012}. Namely, split the Hamiltonian in Eq.~(\ref{Hpf}) into two pieces $H=JV + fF$ in the obvious way, so that $[\alpha_1,V]=0$ (and similarly for $\alpha_{2L}$). Because $[H,\alpha_1]=f\,[F,\alpha_1]$ is non-vanishing, the first step is to find some $X$ satisfying $[{F},\alpha_1]=[V,X]$. If such an $X$ exists, then $[H,\alpha_1 - (f/J)\, X] $ is of order $f^2$.  One can attempt to repeat this procedure to construct strong zero modes as a power series in $f/J$.  This exercise is straightforward for the Ising/Majorana case \cite{Kitaev:2001}; however, one finds \cite{Fendley:2012} for the $\Z3$ parafermion chain that $X$ is proportional to $1/\sin(3\phi)$.  Thus the iteration immediately breaks down for the non-chiral ferromagnetic limit---despite the robust topological bulk gap---implying the demise of strong edge zero modes at arbitrarily small but finite $f/J$.  
Conversely, Ref.~\onlinecite{Fendley:2012} proved that the iterative procedure can be implemented at all orders for $\sin(3\phi)\ne 0$.  Strong edge zero modes therefore exist for sufficiently small $f/[J\sin(3\phi)]$, although an explicit all-orders expression is not known.  As in the Ising/Majorana case, the uniform couplings considered here are not actually necessary; the above arguments still work when $f$ and $J$ vary spatially. 

Perturbation theory in $f$ provides a complementary and illuminating way of understanding this curious result.\cite{Jermyn}  First, we verify that the ground states in the topological phase remain exponentially degenerate for sufficiently small $f/J$, as required for weak parafermion zero modes and hence topological behavior to exist.  For the following it proves convenient to revert back to the equivalent clock-model representation and label the three spin orientations at each site by $A$, $B$, and $C$.   At $f = 0 $ and $|\phi| < \pi/3$ the system exhibits three ground states given simply by $|AAA\dots\rangle$, $|BBB\dots\rangle$, and $|CCC\dots\rangle$.  Since the $f$ perturbation winds single spins, its action on, say, the $A$ ground state gives a sum over excited states of the form $f e^{i\theta}( |BAA\dots\rangle + |ABAA\dots\rangle + \dots) +  f e^{-i\theta} (|CAA\dots\rangle + |ACAA\dots\rangle +\dots)$, all having an excitation energy of order $J$.  The first perturbative correction to the ground-state energies then arises at order $f^2/J$, but crucially, $\Z3$ symmetry requires it be identical for all three ground states.  The energy {\em splitting} between different ground states occurs only at order $L$ in perturbation theory because mixing ground states requires shifting all $L$ spins. This yields a splitting proportional to $(f/J)^L$, indeed exponentially small in system size. Such degeneracy is all that is necessary for the persistence of a topological phase in the parafermion chain.  

Qualitatively different behavior appears for the excitations.  Consider the ferromagnetic $\phi = 0$ limit.  The lowest-lying excitations arise from six flavors of `one-kink' states of the form $|\dots AAABBB\dots\rangle$, $|\dots BBBAAA\dots\rangle$, etc., all exactly degenerate at $f = 0$.  Winding of single spins induced by the $f$ perturbation not only allows the kinks to hop, but in an open chain \emph{also} enables efficient conversion of one flavor to another.   More precisely, sequential actions of $f$ terms can produce the following sequence: $(i)$ An $|\dots AAABBB\dots\rangle$ kink moves to the right end of the system where it becomes $|\dots AAAB\rangle$, $(ii)$ winding the rightmost spin changes kink flavor to $|\dots AAAC\rangle$, and $(iii)$ hopping the kink leftward similarly allows $|ACCC\dots\rangle \rightarrow |BCCC\dots\rangle\rightarrow |\dots BBBCCC\dots\rangle$.  The key is that we connected states related by a global $\mathbb{Z}_3$ transformation---precisely those linked by strong parafermion zero modes at $f = 0$---without leaving the degenerate subspace.  These excitations thus acquire \emph{power-law} splitting at any finite $f$, implying the immediate demotion of strong to weak zero modes.\cite{Jermyn}  

In the chiral case, broken spatial parity causes the energy of the intermediate $|\dots AAACCC\dots\rangle$ kink states to differ from that of the initial $|\dots AAABBB\dots\rangle$ state.  An enormous energy barrier consequently develops that sharply suppresses the above process and drives exponential instead of power-law splitting for the low-lying excitations.  Thus chirality naturally favors the revival of strong zero modes, consistent with the iterative construction above.  

Intriguingly, a sharp transition (tuned, e.g., by the chiral phase $\phi$) appears to separate couplings that produce strong vs.\ weak edge zero modes.  Detailed numerics, for instance, do indeed see abrupt changes between exponential and power-law splitting among excited states.\cite{Jermyn}  Even more interestingly, the numerics suggest a continuous transition, in the sense that the splittings conform well to a scaling ansatz with `critical' exponents.  
A particularly striking feature worth emphasizing is that this transition is purely a property of excited states; indeed the ground states in the topological phase always exhibit exponential splitting and show no signs of `criticality'.  In this sense the physics is reminiscent of transitions discussed in the context of many-body localization \cite{Huse:2013,Bahri:2013}.

\subsection{Parafermions in conformal field theory}
\label{ParafermionCFT}

Conformal field theory enters the physics of topological phases in multiple ways.  It provides a way of classifying all the operators and characterizing their operator products---i.e., their ``fusion"---in many 1D quantum critical points.  Such theories, for instance, describe continuous transitions out of topological phases, including in the parafermion chain. They also describe the gapless edges of many two-dimensional topological phases.  Remarkably, the same underlying mathematical structure remains important even away from quantum criticality.  In fact it also characterizes the fusion and braiding of anyons in the gapped 2D topological phase that we will construct in Sec.~\ref{FibSection} by coupling critical chains together.  

Parafermionic conformal field theory, introduced and described in depth by Zamolodchikov and Fateev \cite{ZamolodchikovParafermion}, governs the scaling limit of the three-state Potts ferromagnetic critical point [$f=J$ and $\theta=\phi=0$ in Eqs.~(\ref{Hclock}) and (\ref{Hpf})] that separates the topological and trivial phases of the parafermion chain; see Fig.~\ref{Chain_fig}(c) for an illustration.  Not surprisingly, there are parafermionic fields---conventionally labeled $\psi_L$ and $\psi_R$---related to the lattice parafermion operators.  The former left-moving field depends on the space-time coordinates $x$ and $t$ solely via the combination $x+vt$ for some velocity $v$, while the latter right-mover depends solely on $x-vt$.   Each exhibits scaling dimension $2/3$. Like their lattice counterparts, they obey the operator products $\psi_L \times \psi_L \sim \psi_L^\dagger$ and $\psi_L^3\sim 1$, and likewise for $\psi_R$. 

More surprisingly, taking the continuum limit of the lattice operator $\alpha_b$ does not give the parafermion fields $\psi_L$ and/or $\psi_R$ as the leading pieces; other more relevant operators mix in as well \cite{Mong:2014,Mong:2014b}.  It is both formally interesting, and essential for the construction of 2D topologically ordered phases reviewed in Sec.~\ref{FibSection}, to find lattice operators that admit exclusively $\psi_L$ or $\psi_R$ as their dominant contribution in the long-distance limit, i.e., to filter out unwanted terms.  Fortunately this is indeed possible.  
A detailed analysis utilizing the problem's discrete symmetries \cite{Mong:2014b} shows that the continuum limit of $\alpha_b$ yields the combination 
\begin{align}
	\alpha_{b} \sim c_1\psi_L + c_2 (-1)^b \Phi\ ,
	\label{alphapsiPhi}
\end{align}
where $c_1$ and $c_2$ denote unimportant numerical constants and $\Phi$ is the unwanted field (which carries dimension $1/15+2/5=7/15<2/3$).  Because of the alternating sign in the last term, a simple linear combination successfully isolates the parafermion field:
\begin{align}
	\alpha_{2a-1}+\alpha_{2a} \sim  \psi_L \ .
	\label{alphapsi}
\end{align}
Finding a lattice analog of $\psi_R$ requires introducing another---though not independent---set of lattice parafermion operators,
\begin{eqnarray}
\beta_{2a-1}&=& \cdots \tau_{a-2}^\dag \tau_{a-1}^\dag (\sigma_a^\dag),
\nonumber \\
\beta_{2a} &=& \cdots \tau_{a-2}^\dag \tau_{a-1}^\dag (\omega^2\tau_a^\dag\sigma_a^\dag) 
\ .
\end{eqnarray}
Compared to Eqs.~(\ref{alphas}) these operators simply have $\tau$ strings running in the opposite direction.  The $\alpha$'s and $\beta$'s are hence related by spatial parity, which immediately implies that
\begin{align}
\beta_{2a-1}+\beta_{2a} \sim  \psi_R \ 
	\label{betapsi}
\end{align}
as desired.

A short overview of the parafermion conformal field theory's fusion algebra, geared toward the present context, appears in Ref.~\onlinecite{Mong:2014}. One feature worth noting is the Fibonacci structure inherent in the  ``energy'' field $E$ of scaling dimension $4/5$. This $\Z3$-invariant field describes the scaling limit of the perturbation that moves the critical parafermion chain into either the topological or trivial phases. In notation introduced in Sec.~\ref{edgezero}, the perturbation corresponds to the lattice operator $F-V$ with $\theta=\phi=0$.  Fusion of the energy field takes the form $E\times E = 1 +E$, which underlies the appearance of `Fibonacci anyons' that we will encounter in Sec.~\ref{FibSection}.

\section{Experimental blueprints for parafermion zero modes}
\label{BlueprintsSec}

We turn now to the crucial question of how one can experimentally realize parafermion zero modes.  In what follows we focus exclusively on weak zero modes, though for brevity we hereafter drop the `weak' qualifier.  Even for this case the problem poses a significant challenge---certainly more so compared to the pursuit of their Majorana cousins.  
For one, the parafermion chain Hamiltonian constructed in Sec.~\ref{LatticeParafermionSec} involves neither bosons nor fermions as elementary constituents; thus at a minimum parafermion zero modes require a strongly correlated host featuring nontrivial emergent degrees of freedom.  The Majorana chain, by contrast, caricatures a $p$-wave superconductor built from electrons and admits a simple free-particle description.  Further subtlety awaits even after we abandon the comforts of free particles: rigorous classifications of 1D bosonic and fermionic systems\cite{FidkowskiKitaev1,Turner:2010,ChenGuWen,Schuch} (subject to a restricted set of symmetries) capture phases supporting Majorana zero modes but nothing more exotic---even with strong interactions!  

Fortunately loopholes exist.  A recurring theme of this section will be the use of `anomalous' edge states in 2D systems as a cradle for 1D phases that host parafermion zero modes.  By definition such edge states can not appear in strictly 1D media and hence fall outside of the aforementioned classifications.  Below we survey several candidate platforms, focusing on those built upon presently available physical systems.  We stress that other interesting, often very closely related proposals exist including an important early work that invokes lattice defects in fractional Chern insulators\cite{Barkeshli:2012}; generalized rotor models\cite{You:2012}; and fractionalized topological insulators and superconductors\cite{Cheng:2012,Vaezi:2013,Klinovaja:2014} (among others).

\subsection{Quantum Hall/superconductor hybrids}
\label{QHSC}

\begin{figure*}[t]
\includegraphics[width=6in]{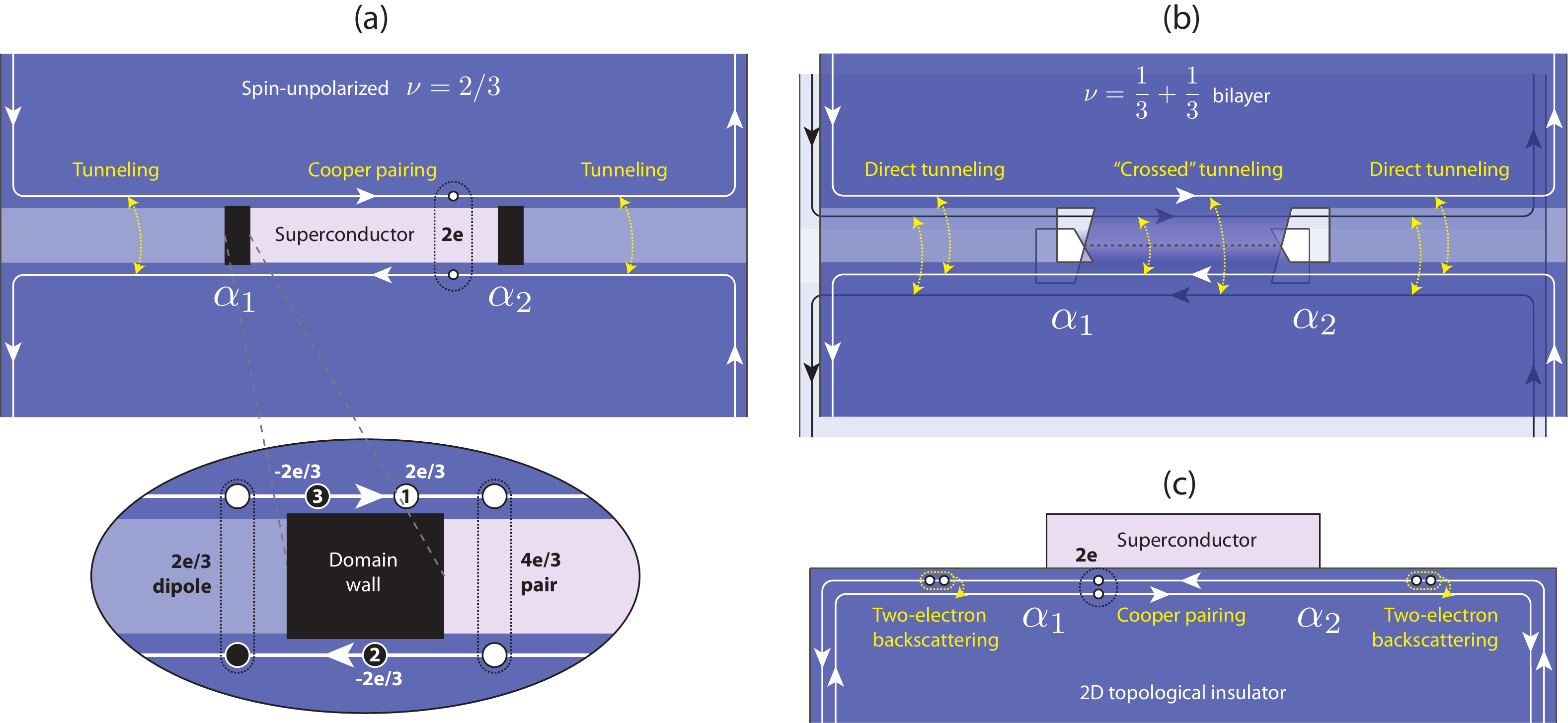}
\caption{Select blueprints for parafermion zero modes $\alpha_{1,2}$.  In (a) only the charged edge mode is shown for simplicity.}
\label{Blueprints_fig}
\end{figure*}

\subsubsection{Setup and physical origin of zero modes}
\label{Setup}

The `anyonic' commutation relation exhibited by the zero modes of interest [recall Eq.~(\ref{paracomm})] spotlights fractionalized 2D systems with anyons as natural parafermion hosts.  Fractional quantum Hall phases provide many experimentally relevant examples, and luckily one can trap parafermions even in the simplest varieties.  For pedagogical purposes we will first examine in some detail a platform that closely parallels various well known Majorana proposals\cite{MajoranaQSHedge,Lutchyn:2010,Oreg:2010}: a spin-unpolarized quantum Hall state at filling $\nu = 2/3$ (i.e., the 112 state) interfaced with an $s$-wave superconductor.  Earlier discussions of other fractions---which closely inform our treatment below---appear in Refs.~\onlinecite{Lindner:2012,Clarke:2013a}.

Figure \ref{Blueprints_fig}(a) sketches the setup.  A trench carved into the $\nu = 2/3$ fluid creates a fractionalized `wire' consisting of edge states that, importantly, support gapless avatars of the anyons living in the adjacent bulk.  Since an identical set of counterpropagating modes appear opposite the trench, physical perturbations can always gap them out.  Electron tunneling across the divide provides one mechanism that simply fuses the two fluids together.  Filling the trench with an ordinary $s$-wave superconductor allows a second, intuitively quite different possibility: generating a gap through proximity-induced spin-singlet Cooper pairing of electrons from the two edges.  These tunneling and pairing mechanisms gap the trench's spin sector in the same way but yield topologically distinct 1D gapped phases for the charge sector.  Domain walls separating the two phases as in Fig.~\ref{Blueprints_fig}(a) bind zero-energy modes that, owing to the system's fractionalization, correspond to $\mathbb{Z}_3$ parafermions.  (For a detailed analysis see Refs.~\onlinecite{Mong:2014,Clarke:2014b}.)

This setup bears the remarkable feature that the spin-singlet Cooper pairs induced by the parent superconductor arise more fundamentally from conglomerates of $2e/3$ charges present in the quantum Hall edges.  The proximity effect therefore not only condenses charge-$2e$ bosons at the trench, but crucially \emph{also} catalyzes `anyon condensation'\cite{You:2013}. To make this precise, we define $\psi_{2e/3}$ to be the spin-conserving quasiparticle operator that adds charge $2e/3$ symmetrically to the trench, with $e/3$ on each end. 
Then in a pairing-gapped region 
\begin{equation}
  \langle\psi_{2e/3}\rangle \sim e^{i \frac{2\pi}{3} p_{2e/3}},
\end{equation} 
where $p_{2e/3}$ takes on integer values. Thus $\langle \psi_{2e/3}^3\rangle$, which essentially yields the expectation value of an ordinary charge-$2e$ Cooper pair, becomes $\sim 1$.
A further remarkable implication follows: although the parent superconductor can only absorb paired electrons, the trench can subsume \emph{fractional} $2e/3$ charges without energy penalty!  

An analogous situation emerges in tunneling-gapped regions. Electron hopping---which also involves multiple fractional charges---catalyzes a different flavor of anyon condensation with 
\begin{equation}
  \langle \psi_{\rm dipole}\rangle \sim e^{i \frac{2\pi}{3} p_{\rm dipole}}, 
\end{equation}
where $\psi_{\rm dipole}$ creates a charge \emph{difference} of $2e/3$ across the trench and $p_{\rm dipole}\in \mathbb{Z}$.  Tunneling-gapped regions accordingly can absorb $e/3$ dipoles with energetic impunity.  Incidentally, this is how tunneling fuses the quantum Hall fluids together; bulk anyons can then move coherently across the trench (leaving behind `invisible' dipoles) even though the intervening region supports only electrons.  

To correctly count the degeneracy encoded by the zero modes, it is essential to observe that the above quasiparticle operators do not commute, 
\begin{equation}
  [\psi_{2e/3}(x),\psi_{\rm dipole}(x')] \neq 0,
\end{equation}
reflecting the fractional statistics of the corresponding bulk anyons.  One therefore can not simultaneously specify a unique phase for both $\langle\psi_{2e/3}\rangle$ and $\langle \psi_{\rm dipole}\rangle$ even though the condensates appear in different spatial regions.  In a basis where the $p_{\rm dipole}$ integers lock to specific values the $p_{2e/3}$ integers fluctuate wildly.  The dipole charges on tunneling-gapped regions correspondingly fluctuate through the allowed low-energy configurations while each paired region realizes degenerate states with well-defined total charge of $0, 2e/3$, or $4e/3$ (mod $2e$).  Conversely, specifying $p_{2e/3}$ requires $p_{\rm dipole}$ to fluctuate, and in this basis the charges on paired regions fluctuate while each tunneling-gapped region takes on one of three degenerate configurations characterized by well-defined dipole charges.  In short, we can label ground states by charge configurations of either the tunneling \emph{or} pairing regions, but not both.  Introducing $2N$ domains walls then yields a degeneracy of $\sqrt{3}^{2N}$ (neglecting Hilbert-space constraints).   It follows that the ground-state degeneracy per domain wall---also known as the quantum dimension---equals $\sqrt{3}$.

The $\mathbb{Z}_3$ parafermion zero-mode operators [denoted $\alpha_{1,2}$ in Fig.~\ref{Blueprints_fig}(a)] simply cycle the system through the ground-state manifold.  To gain further intuition for these modes, consider adding charge $2e/3$ \emph{to one end of the trench}.  Doing so within a pairing-gapped region costs energy (because of the ensuing dipole moment) and similarly for the tunneling regions (due to the net excess charge created).  Within a domain wall, however, the following nontrivial process can uniquely arise:  $(i)$ The added $2e/3$ charge propagates chirally until it hits a superconducting region and Andreev reflects into a $-2e/3$ hole on the opposite side of the trench.  The paired condensate absorbs the deficit $4e/3$ charge.  $(ii)$ The $-2e/3$ hole hops back across the trench upon impinging on a tunneling-gapped region, whose condensate absorbs the resulting change in dipole charge.  Figure \ref{Blueprints_fig}(a) sketches both processes.  $(iii)$ Repeating the preceding steps converts the hole back into a $2e/3$ charge.  Parafermion zero-mode operators essentially create a uniform superposition of these configurations, which costs no energy.\footnote{We thank David Clarke for emphasizing this interpretation.}  This physical picture shows that the zero modes indeed localize to domains walls, add fractional charge of $2e/3$, and thus inherit their nontrivial commutation relations directly from the anyons native to the setup---consistent with the intuition laid out at the beginning of this subsection.  

\begin{figure*}
\includegraphics[width=5in]{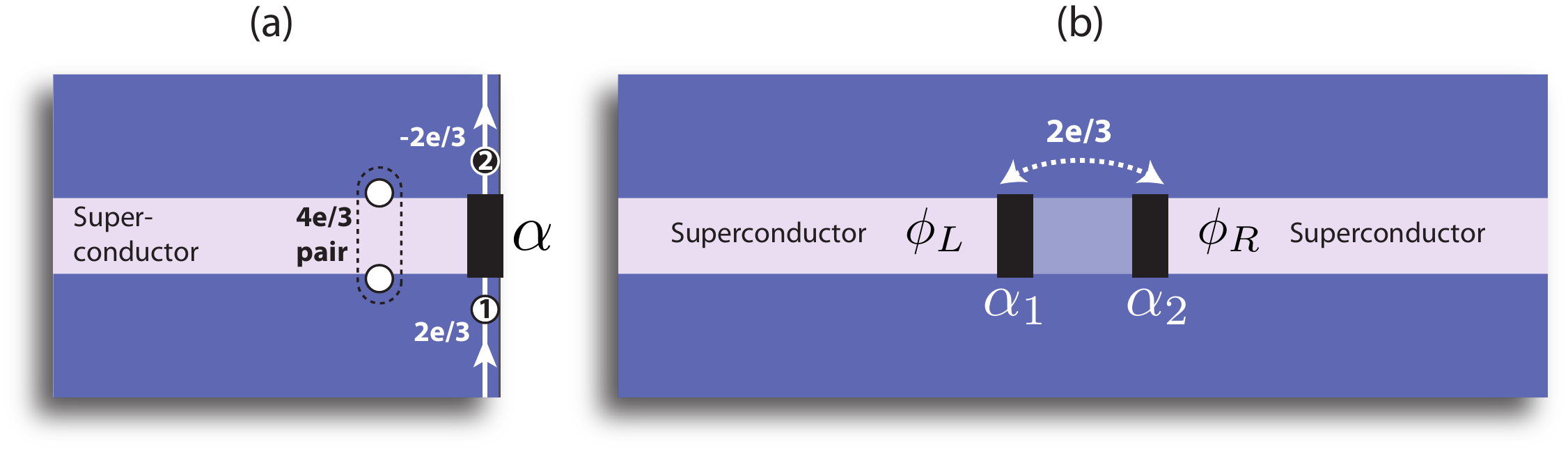}
\caption{Detection schemes for parafermion zero modes, illustrated using the $\nu = 2/3$ setup from Fig.~\ref{Blueprints_fig}(a).  (a) Hybridizing a zero mode with the gapless edge states generates `perfect Andreev conversion'.  (b) Coupling $\mathbb{Z}_3$ parafermions across a Josephson junction yields a current (mediated by $2e/3$ tunneling) $6\pi$-periodic in the phase difference $\phi_L-\phi_R$. }
\label{Detection_fig}
\end{figure*}

\subsubsection{Zero-mode detection}

Armed with these insights, one can anticipate several sharp experimental signatures of parafermions.  Imagine bringing a domain wall to the outer boundary of the host $\nu = 2/3$ quantum Hall fluid as in Fig.~\ref{Detection_fig}(a), thus allowing gapless fractionalized edge states to hybridize with a parafermion zero mode.  In the asymptotic low-energy limit, the zero mode mediates `perfect Andreev conversion':\cite{Clarke:2014b}  Incident $2e/3$ quasiparticles convert with unit probability into $-2e/3$ holes at the domain wall, with the adjacent superconducting region absorbing the deficit $4e/3$ charge.  Two-terminal transport from the edge through the superconductor is thus characterized by a quantized zero-bias conductance $G = 2\nu e^2/h$---\emph{twice} the value obtained when current passes between two normal leads.  Perfect Andreev conversion represents a chiral analogue of the well-studied Majorana-induced perfect Andreev reflection\cite{Sengupta,Law,FidkowskiAlicea,BeenakkerReview2} and provides a conceptually simple fingerprint of parafermion zero modes.  This effect also offers practical utility, as one can exploit its chiral nature to build exotic circuit elements including transistors for fractional charge and flux capacitors.\cite{Clarke:2014b}  

A variant of Fig.~\ref{Detection_fig}(a) further enables detection of the domain wall's quantum dimension through tunneling spectroscopy.  In particular, introducing a second domain wall near the outer edge couples a pair of zero modes, thereby reducing the degeneracy.  The zero-bias conductance peak noted above consequently splits into $g$ finite-voltage resonances\cite{BarkeshliOreg} with $g$ the square of the quantum dimension.  

Consider next Fig.~\ref{Detection_fig}(b) where parafermion modes $\alpha_{1,2}$ hybridize across a Josephson junction with phases $\phi_{L/R}$ on the left/right superconductors.\cite{Clarke:2013a,Lindner:2012,Cheng:2012,ChengLutchyn}  Here the parafermions yield an anomalous Josephson current mediated not by tunneling of ordinary Cooper pairs, but rather fractional $2e/3$ charges.  This effect is actually quite natural given the aforementioned anyon condensation present in the paired regions.  Viewing $e^{i \phi_{L/R}}$ as a charge-$2e$ Cooper-pair creation operator, we see that the anomalous current (in the weak hybridization limit) takes the form
\begin{equation}
  I \propto \sin\left(\frac{\phi_L-\phi_R}{3} + \theta\right),
\end{equation}
where the offset phase $\theta$ depends on the precise state of the barrier.  The Josephson current, in stark contrast to ordinary junctions, thus exhibits $6\pi$ periodicity in the phase difference $\phi_L-\phi_R$!  This phenomenon generalizes Kitaev's $4\pi$-periodic `fractional Josephson effect' for Majorana systems, where supercurrent can arise via single electron tunneling.\cite{Kitaev:2001}  The present setup, however, offers a potential advantage.  Namely, whereas stray unpaired electrons spoil the unconventional periodicity in the Majorana case, the parafermionic Josephson effect is immune to environmental electronic noise since finer fragments carry the anomalous supercurrent.

\subsubsection{Non-Abelian statistics}
\label{BraidingSec}

One can profitably view the superconducting region of the trench in Fig.~\ref{Blueprints_fig}(a) as an extrinsic line defect embedded brute force into the quantum Hall fluid.  The endpoints---where parafermion zero modes localize---represent topological point defects with the key property that a bulk anyon encircling one of the ends swaps the sign of its charge.  [The sign change occurs via an Andreev-like process similar to Fig.~\ref{Detection_fig}(a) when the anyon crosses the paired region.]  Such topological defects are confined since they clearly incur an energy cost that grows linearly with their separation.  Nevertheless it is valuable to view them as `particles', particularly since it is often possible to devise schemes for controllably nucleating and mobilizing parafermion zero modes; see, e.g., Refs.~\onlinecite{Lindner:2012,Clarke:2013a}.  The theory of extrinsic defects in fractionalized media (whose properties intimately relate to symmetries of the topological field theory describing the host platform) has been developed in great detail, e.g., in Refs.~\onlinecite{Bombin,You:2012,Lindner:2012,Clarke:2013a,Cheng:2012,You:2013,Barkeshli:2013a,BarkeshliClassification1,BarkeshliClassification2,Kapustin,TeoRoyChen,TeoRoyChen2,Khan,Wang,Lan,BarkeshliBonderson,TeoHughesFradkin}.  Here we will simply address the following important question: What kind of `particle' is the topological defect that binds a $\mathbb{Z}_3$ parafermion zero mode in our $\nu = 2/3$ setup?  

Suppose that we introduce a set of $2N$ well-separated topological defects into the quantum Hall fluid, generating $\sqrt{3}^{2N}$ ground states (again neglecting Hilbert-space constraints).  Importantly, this degeneracy does not arise from a local property of the defects, consistent with their non-integer quantum dimension, but instead reflects the degenerate global charge configurations on the superconducting regions linking parafermion zero modes.  Hence provided the zero modes maintain their distance no local perturbation can split the ground states; they enjoy topological protection.  Even more interestingly, adiabatically exchanging pairs of defects noncommutatively rotates the system's quantum state within the degenerate manifold.  In other words, \emph{the defects binding parafermion zero modes effectively realize non-Abelian anyons}---a fascinating conclusion given that the host $\nu = 2/3$ system comprises a prototypical Abelian quantum Hall state.  To demystify this seemingly paradoxical statement it is useful to think of the $\nu = 2/3$ state as descending from a parent non-Abelian quantum Hall phase [closely related to SU(2)$_4$]\cite{Hastings:2013,Mong:2014} upon a phase transition that confines the non-Abelian anyons.  By introducing superconducting regions into the fluid, we simply provide the energy necessary to pull apart these confined non-Abelian excitations, whose exotic properties we can still, remarkably, harness.\footnote{We caution, however, that the confinement transition alluded to here can alter the braiding properties of the non-Abelian anyons\cite{Barkeshli:2013a}.}

Widespread interest in non-Abelian anyons stems partly from their potential utility for inherently fault-tolerant topological quantum computing.\cite{Kitaev:2003,Nayak:2008}  The key idea is that the degenerate ground-state manifold non-locally encodes topological qubit states that can be controllably manipulated through defect braiding by virtue of their non-Abelian exchange statistics.  Braiding Majorana zero modes imparts only 90$^\circ$ qubit rotations, thus enabling somewhat restricted fault-tolerant quantum information processing.\cite{Nayak:2008}  It is therefore natural to ask whether our more exotic parafermion zero modes offer comparative advantages.  While in the parafermion case braiding provides additional gates that become trivial with Majoranas,\cite{Clarke:2013a} their exchange statistics remains insufficient for fully universal topological quantum computation.  We note as a teaser that it \emph{is} possible to leverage closely related setups to achieve bona fide computational universality (see Sec.~\ref{FibSection}), but first we explore promising alternative settings for realizing parafermion zero modes.

\subsection{Quantum Hall bilayers}
\label{QHbilayer}

Zooming out slightly, we saw in Sec.~\ref{Setup} that parafermion zero modes arise at the interface between two `incompatibly' gapped regions of a trench in a fractionalized medium.  The Cooper pairing exploited above provides one gapping mechanism that differs fundamentally from ordinary electron backscattering, but interestingly is not always necessary.  Certain quantum Hall phases admit multiple \emph{charge-conserving} gap-opening processes that allow parafermion zero modes to emerge even without superconductivity.  

Figure \ref{Blueprints_fig}(b) depicts an important example consisting of a quantum Hall bilayer with $\nu = 1/3$ filling per layer.\cite{Barkeshli:2013a,Barkeshli:2014b}  A trench through the system can acquire a gap via $(i)$ `direct' tunneling, with electrons backscattering top-to-top and bottom-to-bottom and $(ii)$ `crossed' tunneling where electrons backscatter by moving \emph{between} layers.  Viewing the layer index as a pseudospin, these mechanisms gap the charge sector in the same fashion (inevitably, because charge is conserved) but produce distinct gapped phases in the pseudospin channel.  

Domain walls separating these incompatibly gapped phases once again trap $\mathbb{Z}_3$ parafermion zero modes.  The associated ground-state degeneracy arises because the hopping perturbations catalyze different flavors of anyon condensation that allow the trench to freely absorb `horizontal' intra-layer and `diagonal' inter-layer $e/3$ dipoles in direct- and crossed-tunneling regions, respectively.  Since the operators creating these two dipole types do not commute---again a reflection of fractional statistics of the corresponding bulk anyons---one can label ground states by the charge differences (mod $2e$) on either direct- or crossed-tunneling regions but not both.  The zero-mode operators cycle the system through the degenerate manifold by adding `vertical' $e/3$ dipoles to one side of the trench; such dipoles cost finite energy everywhere except inside of the domain walls, which explains why the parafermions localize.  

As in the superconducting case discussed previously, one can view the endpoints of crossed-tunneling regions as confined topological point defects that carry quantum dimension $\sqrt{3}$ and exhibit non-Abelian statistics.  A bulk anyon winding around such a defect ends up in the opposite layer from which it began (rather than reversing its charge).  Given the quantum dimension, each additional crossed-tunneling region inserted into the $\nu = 1/3+ 1/3$ bilayer increases the ground-state degeneracy threefold.  Curiously, the degeneracy of a \emph{single-layer} $\nu = 1/3$ fluid increases by precisely the same factor when system's genus rises.  This is no coincidence---upon rotating the top layer in Fig.~\ref{Blueprints_fig}(b) by $180^\circ$ about the trench, the crossed-tunneling region locally resembles a puncture drilled into in a $\nu = 1/3$ system on a sphere.  Barkeshli et al.\cite{Barkeshli:2013a} thus christened the topological defects in this bilayer setup `genons'.  

Parafermion zero modes bound to `genons' yield transport fingerprints similar to those arising in quantum Hall/superconductor heterostructure.\cite{Barkeshli:2014b,BarkeshliOreg}  For instance, hybridizing a zero mode with gapless edge states native to the bilayer [in a straightforward generalization of Fig.~\ref{Detection_fig}(a)] generates a perfect-Andreev-reflection analogue where incoming `vertical' dipoles from the edge switch polarity with unit probability.\cite{Barkeshli:2014b}  (The adjacent crossed-tunneling condensate readily absorbs the change in dipole moment.)  More broadly, the similarity between the two setups discussed so far is quite remarkable given the total absence superconductivity in the present case.  While the bilayer architecture nevertheless raises nontrivial fabrication challenges given the somewhat unnatural crossed-tunneling terms that we invoked, Ref.~\onlinecite{Barkeshli:2014b} suggests several plausible experimental routes to realizing the couplings (interestingly, not all of which are essential for the topological degeneracy).

\subsection{2D topological insulator edges}
\label{QSH}

While we have now seen that fractionalized 2D systems with anyons indeed comprise natural parafermion-zero-mode hosts, fractionalization turns out, quite remarkably, to be inessential.  Two important works\cite{ZhangKane,Orth} recently showed that one can instead trap parafermions using a much simpler medium: a 2D topological insulator.  As background, a 2D topological insulator edge supports counterpropagating electron modes described by operators $f_{R/L}$ that form Kramers partners.  These modes necessarily remains gapless provided the system preserves both particle-conservation and time-reversal symmetries.  Conversely, adding superconductivity induces a pairing gap while introducing magnetism can gap the edge via backscattering.  As shown earlier by Fu and Kane\cite{MajoranaQSHedge} domain walls separating these clearly distinct gapped phases bind Majorana zero modes.  More interestingly, replacing the magnetic backscattering terms with strong interactions that \emph{spontaneously} break time-reversal symmetry promotes these Majorana modes to more exotic $\mathbb{Z}_4$ parafermions!

We can capture this result using a bosonized framework where the electron operators and density respectively become $f_{R/L} \sim e^{i(\varphi\pm \theta)}$ and $\rho = \partial_x\theta/\pi$.  The bosonized fields obey $[\varphi(x),\theta(x')] = i\pi \Theta(x-x')$ with $\Theta(x)$ the Heaviside function and $x$ an edge coordinate.  Let us examine in this language the gapping mechanisms available in the absence of explicit time-reversal breaking:  $(i)$ Proximity-induced Cooper pairing with strength $\Delta$ yields a perturbation
\begin{equation}
  H_{\rm pair} = -\int_x(\Delta f_R f_L + H.c.) \sim - \int_x \Delta \cos(2\varphi)
  \label{Hpair}
\end{equation}
that gaps the edge by locking $\varphi = \pi p_{\rm pair}$ with $p_{\rm pair} \in \mathbb{Z}$ to minimize the energy.  
$(ii)$ Time reversal forbids elastic single-electron backscattering but allows two-particle backscattering described by a Hamiltonian
\begin{equation}
  H_{\rm bs} = -\int_x (\Gamma f_R^\dagger f_R^\dagger f_L f_L + H.c.) \sim -\int_x\Gamma \cos(4\theta).
  \label{Hbs}
\end{equation}
With sufficiently strong density-density repulsion the coupling $\Gamma$ becomes relevant and gaps the edge by pinning $\theta = \frac{\pi}{2} p_{\rm bs}$  where $p_{\rm bs} \in \mathbb{Z}$.\cite{XuMoore,ZhangKane,Orth}  The system then spontaneously breaks time reversal symmetry, as indicated by the local magnetization order parameter $\langle f_R^\dagger f_L\rangle \sim (-1)^{p_{\rm bs}}$.  Contrary to the situation with explicit symmetry breaking, however, configurations with opposite magnetization (i.e., even vs.~odd $p_{\rm bs}$) remain degenerate.  This feature proves essential in what follows.  

Consider the setup of Fig.~\ref{Blueprints_fig}(c) where regions gapped by two-electron backscattering straddle a Cooper-paired edge segment proximate to a superconductor.\footnote{Technically, maintaining relevance of both backscattering and pairing terms requires different interaction strengths in the corresponding regions, which is quite reasonable given the screening provided locally by the parent superconductor\cite{Orth}}  The backscattering terms pin $\theta = \frac{\pi}{2} p_{\rm bs}^{l/r}$ on the left/right regions while $\varphi = \pi p_{\rm pair}$ on the superconducting section.  Similar to the situation encountered in Sec.~\ref{Setup}, we can not uniquely fix all of these integers since $\theta$ and $\varphi$ do not commute.  The special physics of this setup becomes most apparent in a basis where we specify $p_{\rm bs}^{l/r}$, leaving the $p_{\rm pair}$ integers wildly fluctuating.   The total charge (mod $2e$) on the central paired segment then reads
\begin{equation}
  Q = e\int_{x_l}^{x_r} dx\frac{\partial_x\theta}{\pi} = e\frac{\theta(x_r)-\theta(x_l)}{\pi} = \frac{e}{2}(p_{\rm bs}^r-p_{\rm bs}^l),
\end{equation}
with $x_{l/r}$ positions just to the left/right of the superconductor.  Hence the superconducting part of the edge admits \emph{four} degenerate states labeled by total charges $0, e/2, e,$ and $3e/2$ (mod $2e$).  Furthermore, we see that raising $p_{\rm bs}^r$ or $p_{\rm bs}^l$ by one unit---thereby reversing the magnetization in an adjacent backscattering region---pumps charge $\pm e/2$ into the superconductor.  Accordingly, localized $\mathbb{Z}_4$ parafermion zero-mode operators cycle the system through the degenerate manifold by adding spin and charge $e/2$ to the domain walls.\footnote{Formally, these operators arise from projection of $e^{i(\varphi/2+\theta)}$, which cycles the fields among adjacent minima of the cosines in Eqs.~(\ref{Hpair}) and (\ref{Hbs})}  

Hybridizing a pair of $\mathbb{Z}_4$ parafermions across a Josephson junction [similar to Fig.~\ref{Detection_fig}(b)] yields an anomalous $8\pi$-periodic Josephson current reflecting $e/2$ transport across the barrier.  Perhaps the most remarkable feature of this setup is that while isolated parafermion zero modes require strong electron-electron repulsion---a nontrivial scenario to obtain in practice---\emph{the $8\pi$-periodic Josephson effect can survive even with arbitrarily weak interactions}.\cite{ZhangKane}  This beautiful result may be understood by adiabatic continuity: Starting from the strongly coupled limit, dialing down the barrier's interaction strength yields only smooth crossover behavior due to finite-size effects, hence the robustness of the $8\pi$ periodicity.  

Since trapping parafermion zero modes does not require systems with anyons, one might hope that even simpler purely 1D wires suffice.  Several works have explored this intriguing issue\cite{Oreg:2014,Klinovaja:2014a,Klinovaja:2014b,Tsvelik:2014b}, providing possible routes that do, however, require some amount of fine-tuning---in accord with the 1D classifications discussed at the start of Sec.~\ref{BlueprintsSec}.  If parafermion zero modes can stably appear in strictly 1D setups they will likely require symmetry protection; finding explicit examples (or proving otherwise) remains an important open question.

\section{Fibonacci anyons from coupled parafermions}
\label{FibSection}

\begin{figure*}
\includegraphics[width=5in]{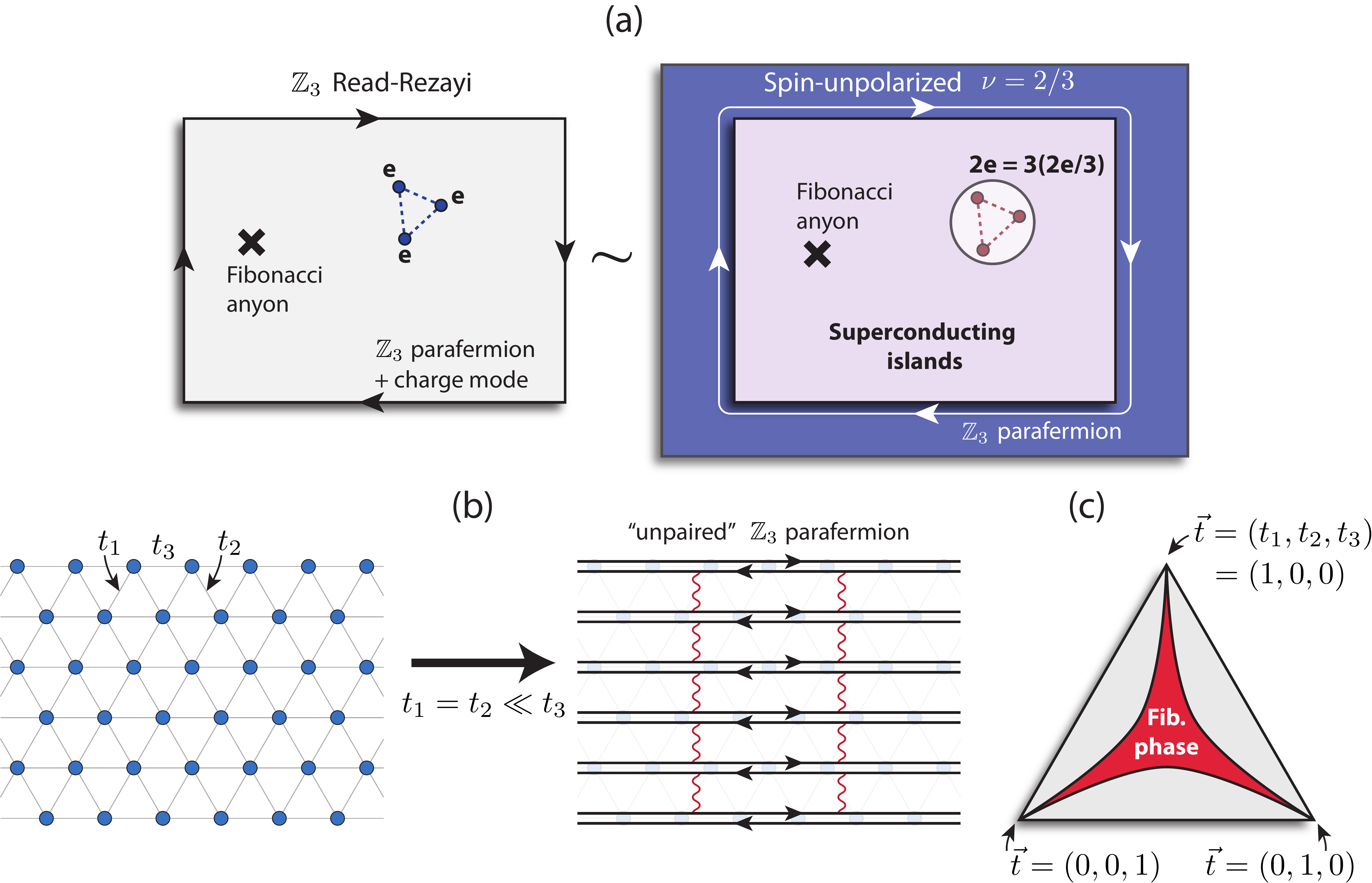}
\caption{(a) Correspondence between the $\mathbb{Z}_3$ Read-Rezayi state and the `Fibonacci phase' formed in a $\nu = 2/3$ quantum Hall/superconductor heterostructure.  (b) Triangular-lattice parafermion model (left) that realizes the Fibonacci phase in the weakly-coupled-chain limit (right).  (c) Phase diagram of the setup in (b) obtained from the DMRG analysis of Ref.~\onlinecite{Stoudenmire:2015}.}
\label{Fib_fig}
\end{figure*}

Braiding topological defects binding parafermion zero modes yields denser qubit rotations compared to the Majorana case, but not sufficiently dense to enable computational universality.  In search of universal topological quantum computing hardware we must therefore look to still more exotic non-Abelian excitations.  Following the spirit of the above analysis, here we discuss a way to ``engineer'' the requisite topological order.

For inspiration, we recall that potential ``intrinsic'' sources may arise from highly exotic quantum Hall phases.  The $\mathbb{Z}_3$ Read-Rezayi state\cite{Read:1999}, a natural generalization of the famed Moore-Read state\cite{Moore:1991}, provides a notable example.  This phase features a gapless edge structure consisting of a charge mode together with a neutral sector described by a chiral $\mathbb{Z}_3$ parafermion conformal field theory (which comprises half of the non-chiral theory discussed in Sec.~\ref{ParafermionCFT}).  As an intimately related consequence the bulk supports non-Abelian `Fibonacci anyons' with the following properties: their quantum dimension is the golden mean, $\varphi = (1+\sqrt{5})/2$; two Fibonacci anyons brought together  either annihilate or beget another Fibonacci anyon; and, most interestingly, their braiding provides a universal gate set\cite{Freedman:2002a,Freedman:2002b}.  
Fibonacci anyons thus constitute a holy grail for topological quantum computation.  Ultimately the special physics of the Read-Rezayi state derives from a peculiar property built into the ground state: clustering correlations amongst \emph{triplets} of electrons (as opposed to pairs that occur in a superconductor).  

The comparatively simple Abelian quantum Hall systems exploited to obtain $\mathbb{Z}_3$ parafermion zero modes in Sec.~\ref{BlueprintsSec} can, remarkably, provide close cousins of the Read-Rezayi state.  As a concrete example, imagine patterning a 2D array of superconducting islands into a $\nu = 2/3$ state, thereby generating a network of $\mathbb{Z}_3$ zero modes that encode a massive ground-state degeneracy reminiscent of the situation in a partially filled Landau level.  Judiciously hybridizing adjacent parafermion zero modes can then lift the degeneracy in favor of a superconducting `Fibonacci phase' hosting a gapless chiral $\mathbb{Z}_3$ parafermion mode at its boundary and Fibonacci anyons in the interior.\cite{Mong:2014}  This outcome is actually quite intuitive.  As emphasized previously, the superconducting islands impart condensation of charge-$2e$ Cooper pairs that derive from conglomerates of three fractionalized $2e/3$ quasiparticles in the quantum Hall fluid.  Thus the heterostructure effectively leverages pairing to mimic multi-particle clustering---making Read-Rezayi-like physics rather natural.  The resulting correspondence between the superconducting Fibonacci phase and the Read-Rezayi state [summarized in Fig.~\ref{Fib_fig}(a)] generalizes Read and Green's\cite{Read:2000} classic work demonstrating the common non-Abelian content of a $p+ip$ superconductor and the Moore-Read state.  

Analytical and numerical analyses substantiate the preceding physical argument. One powerful method assembles the 2D Fibonacci phase from coupled chains similar to Teo and Kane's seminal work\cite{Teo:2014} that obtained non-Abelian quantum Hall phases from Luttinger liquid arrays.  Here, however, we utilize the parafermion chains described in Sec.~\ref{sec:chains} instead of Luttinger liquids. Namely, we consider $\mathbb{Z}_3$ parafermion operators arranged on a triangular lattice with nearest-neighbor couplings $t_1, t_2$, and $t_3$ sketched in Fig.~\ref{Fib_fig}(b) \cite{Stoudenmire:2015}.  The setup could arise either in a $\nu = 2/3$ state with superconductivity or a $\nu = 1/3+1/3$ bilayer with crossed-tunneling defects, among others.  

In the extreme limit $t_{1,2} = 0$ the problem reduces to decoupled Potts chains of Sec.~\ref{LatticeParafermionSec}, with $J = f = t_3$ and $\phi = \theta = 0$.  As Fig.~\ref{Chain_fig}(c) illustrates each chain therefore realizes a non-chiral $\mathbb{Z}_3$ parafermion critical point.  The system's gaplessness here is a great virtue since even exceedingly weak interchain couplings $t_{1,2}$ can profoundly impact the physics.  
Crucially, the relation between lattice parafermion operators and fields in the corresponding conformal field theory is known\cite{Mong:2014b}; recall Eqs.~(\ref{alphapsi}) and (\ref{betapsi}).  This relation allows a systematic analysis of the weakly-coupled-chain limit that `filters out' irrelevant lattice-scale details by projecting onto low-energy fields describing long-wavelength physics.  (Reference~\onlinecite{VaeziFib} pursues a similar strategy, though we caution that their construction utilizes unphysical interchain couplings; see Ref.~\onlinecite{Mong:2014b} for further discussion.)  

An especially desirable limit arises when $t_1 = t_2\equiv t$, where the most relevant interchain perturbation becomes simply\cite{Mong:2014,Stoudenmire:2015}
\begin{equation}
  \delta H \sim -t \sum_y \int_x (\psi_{R,y}^\dagger \psi_{L,y+1} + \hc),
\end{equation}
with $\psi_{L/R,y}$ parafermion fields for chain $y$.  The wavy lines on the right side of Fig.~\ref{Fib_fig}(b) designate the pattern of couplings, which completely gap the interior but leave behind `unpaired' \emph{chiral} $\mathbb{Z}_3$ parafermion modes at the boundary. Such a perturbation of the parafermion critical point in purely 1D field theory (i.e., if we instead coupled $\psi_R$ to $\psi_L$ in the \emph{same} chain), results in a Fibonacci-type spectrum, as known exactly from integrability\cite{Fateev:1990}.  In our coupled-chain system, the perturbation correspondly triggers precisely the Fibonacci phase anticipated above.  While this analytic approach breaks down at stronger interchain couplings, density matrix renormalization group (DMRG) simulations allow a controlled exploration of the broader parameter space.\cite{Stoudenmire:2015}  Figure~\ref{Fib_fig}(c) sketches the numerically determined phase diagram; fortunately the Fibonacci phase enjoys a wide stability window extending from weakly coupled chains to the isotropic triangular lattice limit ($t_1 = t_2 = t_3$) and beyond! 

It is quite encouraging that a non-Abelian state suitable for universal topological quantum computation can in principle emerge upon combining well understood components such as a $\nu = 2/3$ state and conventional superconductors.  Similar phases have also been predicted in various related settings including uniform quantum Hall bilayers\cite{Vaezi:2014}, quantum wire\cite{Sagi:2014} and spin chain\cite{Meng} arrays, and even local bosonic models\cite{BarkeshliFib}. We also note that an interesting earlier study, Ref.~\onlinecite{Burrello:2013}, explored other coupled 2D parafermion lattice models and captured an Abelian topologically ordered phase that generalizes the toric code. These developments raise a number of intriguing questions: How can the architectures reviewed here be further distilled to ease practical implementation?  What probes enable unambiguous detection of Fibonacci anyons in such setups?  How might they be manipulated to reveal their coveted braiding statistics?  And can one pursue similar philosophies to obtain perhaps even more exotic non-Abelian phases?

\section{Outlook}
\label{OutlookSection}

Parafermions boast a more than three-decade history in condensed matter, with numerous now-classic applications ranging from criticality in Potts chains to the non-Abelian Read-Rezayi quantum Hall phases.  Despite this lengthy history, understanding of parafermions---both on a formal and practical level---has advanced significantly during the past few years.  Much of this progress can be viewed as extending the wildly influential Ising-model physics (in its myriad of incarnations) into more exotic territory.  

On the formal end we saw in Sec.~\ref{sec:chains} that para\-ferm\-ion chains exhibit a topological phase supporting parafermion zero modes that comprise natural Majorana generalizations.  While appealing parallels with the simpler Ising/Majorana chain do exist, we encountered subtle surprises owing to the inherently strongly interacting nature of parafermions.  As one intriguing example, even deep within the topological phase strong parafermion zero modes (which guarantee threefold degeneracy of \emph{all} eigenstates) can abruptly transform into weak zero modes due to an unusual form of quantum criticality operating solely within the excited states.  The effect of disorder on this transition raises an interesting issue that warrants further exploration; in particular, following Refs.~\onlinecite{Huse:2013,Bahri:2013} it is tempting to anticipate a many-body localized phase that resurrects strong zero modes even when they are absent in the clean system.  

Another recent formal advance established the precise correspondence between lattice parafermion \emph{operators} and continuum parafermion \emph{fields} emerging in the critical chain.  These relations not only clarify certain aspects of the conformal fields theory, but also underlie controlled analytical techniques that attack 2D systems built from parafermion chains.  Yet another example is the identification of a `free-parafermion' limit obtained in a deformed non-Hermitian parafermion chain model.\cite{Fendley:2014}  Might it be possible to bootstrap off of this limit to controllably attack strongly interacting parafermion systems?  Or can one perhaps design implementations of the free-parafermion model itself?  

By now many proposed experimental realizations of parafermion zero modes exist.  We focused in Sec.~\ref{BlueprintsSec} on three classes of systems that appear especially promising in that they rely entirely on well-understood phases of matter: Abelian quantum Hall/superconductor hybrids, simple quantum Hall bilayers with `genon' defects, and strongly interacting 2D topological insulator edges coupled to superconductivity.  The first class perhaps appears a somewhat awkward marriage given the high fields usually required to observe fractional quantum Hall physics.  We stress, however, that very recent experiments successfully induced a proximity effect in GaAs quantum wells using high-critical-field ($>10$T) NbN superconducting contacts.\cite{Wan}  This development bodes well for the prospects of exploring the interplay between quantum Hall physics and superconductivity---a largely untapped experimental arena ripe for discovery.  
Bilayer `genon' proposals require controlled manipulation of edge states to selectively hybridize different pairs of counterpropagating modes, which although nontrivial seems achievable with presently available gating technology.\cite{Barkeshli:2014b}  Such systems carry the added bonus that detecting parafermion zero modes can also reveal (albeit indirectly) the topological ground state degeneracy of quantum Hall phases on nontrivial-genus manifolds.\cite{BarkeshliOreg}  
Interacting 2D topological insulator edges may offer the most immediate prospects of detecting signatures of parafermion zero modes.  Two virtues are particularly noteworthy here: First, experiments have already achieved the required coupling to superconductors,\cite{Hart,Pribiag} and second, the anomalous $8\pi$-periodic Josephson effect that parafermions mediate survives even in the weakly interacting limit.\cite{ZhangKane}  

One can view arrays of coupled parafermion zero modes in such setups as new low-energy degrees of freedom from which even more exotic higher-dimensional topological phases can emerge.  Our understanding of the possibilities is certainly far from complete, though initial studies have already revealed exceptionally rich physics.  Notably, networks of parafermion modes arising in the above quantum-Hall-based platforms can yield a `Fibonacci phase' that represents an analogue of the coveted $\mathbb{Z}_3$ Read-Rezayi state with all but the most interesting non-Abelian Fibonacci anyons---which carry universal braid statistics---distilled away.  This insight provides proof of principle that well understood components can combine to yield universal topological quantum computing hardware.  Simplifying the requisite architectures poses an important outstanding challenge.  One promising avenue is to wield numerics to examine the phase diagrams of \emph{uniform} quantum Hall/superconductor hybrids and quantum Hall bilayers in search of related non-Abelian phases; studies in the latter setting have in fact recently been carried out\cite{Geraedts,Peterson,Liu}.

While experimental prospects and tantalizing long-term technological applications are certainly exciting, in our view it is worth emphasizing the inherent theoretical beauty underlying the advances in parafermions surveyed here.  The physics that has so far emerged intimately links seemingly disparate areas across condensed-matter physics to an extent that rivals the reach of the Ising model.   As this story continues to evolve, undoubtedly many more fundamental insights will be gained in this fascinating field.

\section*{Acknowledgments}
We are indebted to D.~Aasen, E.~Berg, D.~Clarke, A.~Essin, M.~P.~A.~Fisher, A.~Jermyn, N.~Lindner, R.~Mong, D.~Mross, C.~Nayak, Y.~Oreg, K.~Shtengel, A.~Stern, and M.~Stoudenmire for collaborations related to this review.  
We also gratefully acknowledge support from the National Science Foundation through grant DMR-1341822 (J.\ A.); the Alfred P.\ Sloan Foundation (J.\ A.); the Caltech Institute for Quantum Information and Matter, an NSF Physics Frontiers Center with support of the Gordon and Betty Moore Foundation through Grant GBMF1250; and the Walter Burke Institute for Theoretical Physics at Caltech.


%

\end{document}